\documentclass{article}
\usepackage[utf8]{inputenc}
\usepackage{graphicx}
\usepackage{caption}
\usepackage{amsmath,pdftexcmds}
\usepackage[a4paper, total={6.5in, 10in}]{geometry}
\usepackage{color}
\usepackage{braket}
\usepackage{comment}
\usepackage{abstract}
\usepackage{amsfonts}
\usepackage{subfig}
\usepackage{graphicx}
\usepackage{multirow}
\usepackage{caption}
\usepackage{pdfpages}
\usepackage{amssymb}
\usepackage{float}
\usepackage{pgfplots}
\usepackage{tikz}
\usepackage{tikz-3dplot}
\usepackage{authblk}
\usepackage[none]{hyphenat}
\usepackage{epsfig}
\graphicspath{graphics/}
\usepackage[colorlinks,linkcolor={black},citecolor={black},urlcolor={blue},unicode,breaklinks=true]{hyperref}

\title{\textbf{Supercoiling DNA with a free end}}

\author{Daniela Moretti}
\author{Giuseppe Gonnella}
\author{Antonio Suma}
\affil{Dipartimento Interateneo di Fisica, Universit\`a degli Studi di Bari and INFN, Sezione di Bari, via Amendola 173, Bari, I-70126, Italy. E-mail: daniela.moretti@uniba.it.}
\author{Giada Forte}
\author{Davide Marenduzzo}
\affil{SUPA, School of Physics and Astronomy, University of Edinburgh, Edinburgh EH9 3FD, United Kingdom.}
\author{Cristian Micheletti}
\affil{Scuola Internazionale Superiore di Studi Avanzati - via Bonomea, 265 - 34136 Trieste, Italy.}

\date{}

\pgfplotsset{compat=1.18}
 
\begin{document}

\maketitle

\begin{abstract}
In this work, we combine coarse-grained Brownian dynamics simulations and mean-field theory to study supercoiling dynamics, as well as the steady-state profiles of twist and writhe, in an open DNA polymer where one of the free ends is subjected to a constant torque. Even though the other end is free, and hence can spin and release torsional stress, we observe that the entire chain transitions between a swollen and a plectonemic phase as the torque increases beyond a critical threshold. 
In the plectonemic phase, we observe a non-linear twist profile in the steady state, resulting from the mutual interconversion between the injected twist and geometrical writhe, which distributes inhomogeneously along the chain. We also show that the non-equilibrium dynamics of twist accumulation is diffusive, and that writhe diffusion is negligible in this geometry, as plectonemes remain localised near the end that is being rotated. We discuss the feasibility of testing our results with single-molecule experiments. 
\end{abstract}

\section{Introduction}

DNA supercoiling refers to the under- or over-twisting of the DNA duplex, which in equilibrium in its most stable B-form is a right-handed double helix, with a pitch of about $10.5$ base pairs, or $3.5$ nm~\cite{Calladine1997,Bates2005}. Supercoiling is highly relevant biologically because it is deeply entwined with function. On one hand, transcription naturally introduces supercoiling: according to the ``twin supercoiling domain'' model~\cite{Liu1987,Mielke2004,Fosado2021,Joyeux2024} an active RNA polymerase, whose rotation is hindered, generates a wave of positive supercoiling ahead of it (downstream), and one of negative supercoiling behind (upstream). Transcriptionally-induced supercoiling provides a mechanistic pathway to promote gene bursting~\cite{Chong2014,Ancona2019}. It may be beneficial because negative supercoiling can facilitate the recruiting of RNA polymerase after a gene has been transcribed, and thus trigger a positive feedback loop that upregulates promoters~\cite{Brackley2016, Sevier2017}.  

Supercoiling in double-stranded DNA can manifest itself as either twist or writhe~\cite{Bates2005,Marko1994,Marko1995}.  For a torsionally constrained molecule whose ends are fixed or held under tension, changes in linking number—the topological invariant measuring how many times the two strands wind around each other—can induce a buckling transition between a stretched phase, in which the excess linking is stored mainly as twist, and a plectonemic phase, in which it is stored mainly as writhe~\cite{Marko1994,Marko1995,Marko1997,Marko2007}. Fixed boundary conditions are crucial in this context, as the DNA linking number is strictly conserved only if twist and writhe cannot escape through either nicks of the duplex or free ends. The buckling transition can be assimilated to a gas-to-liquid transition~\cite{Marko2007}, where the stretched and plectonemic phases play the roles of the gas and liquid phases, respectively. The conservation of the linking number then leads to model B-like dynamics~\cite{Chaikin1995} and phase separation between stretched and plectonemic domains~\cite{Forte2019}. Within this framework, the typical plectoneme size at equilibrium is entirely controlled by thermal fluctuations~\cite{Skoruppa2022}. 

Here we instead consider an open  DNA filament with one {\it free end}, and use coarse-grained Brownian dynamics simulations~\cite{Chirico1994,Brackley2014} to study the response of the duplex to the application of a constant torque at the other end. We ask whether the continuous injection of twist can still drive a transition to a plectonemic phase despite the fact that the free end can swivel. Different from the buckling case, such a transition would be inherently out of equilibrium. Our setup is related to those considered in~\cite{Wada2009,Laleman2016}, but differs in several ways. In Ref.~\cite{Wada2009}, twist is injected at one end via a constant angular velocity, and hydrodynamic interactions are included for studying plectoneme friction in short DNA chains. Ref.~\cite{Laleman2016} studied the morphology of a twistable polymer constrained to wrap around a straight rod. Our study complements these previous ones for both the setup and the goals. We use Langevin dynamics to study relatively long twistable polymers, parametrized after DNA, that can freely fluctuate in three dimensions while a constant torque is applied at one of the ends. For simplicity, hydrodynamic interactions are neglected.
 In addition, we focus on the build-up, spatial distribution and steady-state profile of supercoiling (twist and writhe) along the chain.

This free boundary problem is much less studied than its fixed boundary or stretched end counterparts, but it could, in principle, be realised in single molecule experiments.
In fact, torque can be applied either via RNA polymerases~\cite{Ma2014,Janissen2024}, optical or magnetic tweezers~\cite{Strick1996,gao2021torsional}, or  via electrosmotic flow~\cite{zheng2025torsion,maffeo2023dna}, while imaging of resulting plectonemes can be detected via fluorescence imaging~\cite{Janissen2024}.
It may also be relevant to model {\it in vivo} genomic segments that are far from topological barriers that hinder supercoiling propagation. Additionally, from a purely theoretical perspective, it provides a natural setting for studying transport and diffusion of supercoiling~\cite{Nelson1999,Brackley2016,Fosado2021,Fosado2021b}, as well as the competition between twist injection and supercoil escape from the polymer ends. The latter process can  supposedly occur via multiple modes, making the efficiency of supercoil release from the free end a non-trivial dynamical question. In the first mode, the free end is assumed to spin without resistance and thus efficiently dissipate injected twist~\cite{Levinthal1956}. However, this modality can be considered the sole one only if the chain remains approximately cylindrical, does not bend and has a sufficiently high torsional rigidity ~\cite{Nelson1999}; otherwise, if writhe and plectonemes  develop along the chain, one might expect injected twist to be dissipated either via solid-body rotation, or via plectoneme  diffusion past the chain end\cite{Wada2009}, with a dynamics that is expected to be  much slower compared to pure twist diffusion~\cite{Fosado2021}.
 
We find that, even without long-range hydrodynamic interactions, increasing the applied constant torque past a threshold value leads to a transition between a swollen and a plectonemic phase, similar to what observed by~\cite{Wada2009} for the constant rotational velocity ensemble. Unlike the conventional buckling transition~\cite{Neukirch2011}, where the ends are under tension and the supercoiling density is controlled at fixed linking number, here the torque is maintained constant and thus twist is continuously injected, driving the polymer into a non-equilibrium steady state where it has a finite rotational velocity. Plectoneme formation, therefore, results from a different balance with respect to the usual stretching ensemble. In the non-equilibrium steady states we obtain, we identify a characteristic regime where plectoneme formation leads to a non-linear twist profile in steady state. To rationalize this, we introduce a mean field argument for the coupled evolution of local twist and writhe in the presence of a free end. Finally, the evolution of plectonemes is also strongly affected by the free-end setup, which favours their nucleation at the rotating (torque-driven) boundary, where they tend to remain pinned throughout the simulations. 

Our work is structured as follows. In Section \ref{model}, we present the twistable polymer model, along with details about its implementation in LAMMPS. 
In Section~\ref{linear_polymer}, we overview the  phenomenology of the model in the limiting case of a perfectly straight polymer chain, which provides a baseline for twist transport.
In Section~\ref{results}, we present the main results for a initially equilibrated configurations of the model DNA duplex. 
Finally, we conclude by summarizing our findings and discussing possible refinements of the model and experimental validations of the predictions.

\section{The polymer model}\label{model}
\begin{figure*}
    \centering
    \includegraphics[width=0.95\linewidth]{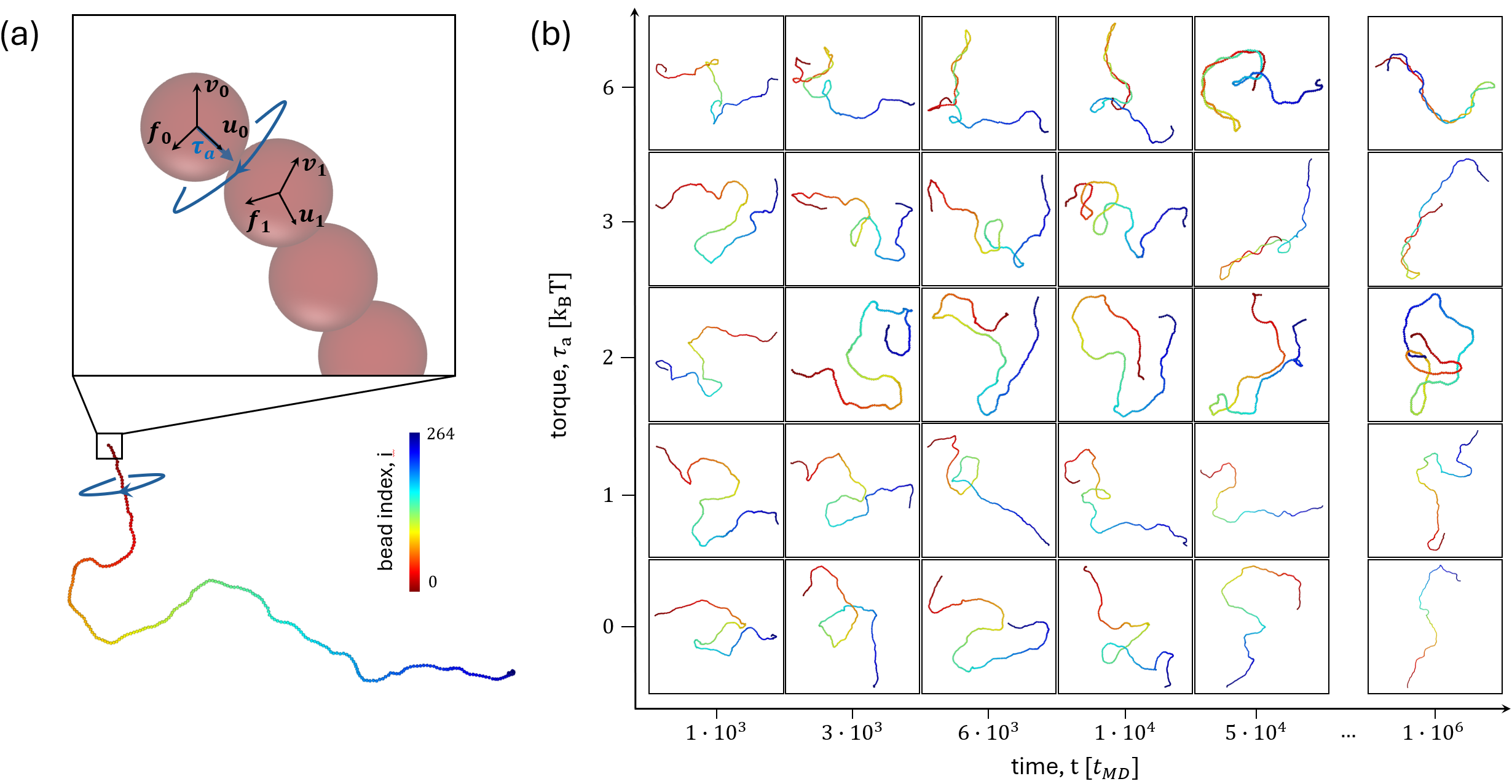}
    \caption{(a) Initial conformation at $t=0$ of a $2 {\rm kbp}$ polymer. The inset shows an enlargement of the first bead index, with the orthonormal set $\{\mathbf{f}_i, \mathbf{v}_i, \mathbf{u}_i\}$ sketched for the first two beads. Highlighted is the active torque vector $\tau_a$, which is applied only on the bead $i=0$ in the direction of $\mathbf{u}_0$ and approximately parallel to the tangential vector $\mathbf{t}_0$. The circles highlight the direction of rotation caused by $\tau_a$. (b) Snapshots of  $2 {\rm kbp}$ long polymers subject to increasing torques $\tau_a=0,1,2,3,6$ (from bottom to top), and for increasing simulation times  (from left to right). Chains are colored by chain index, see color bar in (a). Supplementary Movies S1-S5 display the complete trajectories for the cases illustrated in (b).}
    \label{fig:configurations}
\end{figure*}

To describe the dynamics of double-stranded DNA, we employ a coarse-grained bead-chain model within the framework of Langevin dynamics, following the approach initially proposed by Chirico and Langowski~\cite{Chirico1994}. 
Each bead $i$ has a diameter $\sigma=2.5$~nm, corresponding to approximately 7.3 base pairs (bp), and the chain consists of $N$ beads in total, with the bead index $i$ running
from 0 to $N-1$. Bead $i$ is characterized by its Cartesian position $\mathbf{r}_i$ and an orthonormal triad of unit vectors $\{\mathbf{f}_i, \mathbf{v}_i, \mathbf{u}_i\}$ (see Fig.~\ref{fig:configurations}(a)) defining the local body frame. The frame allows us to account for bending, through changes of $\mathbf{u}_i$ along the contour,  and twist, through rotations around $\mathbf{u}_i$.
To mimic the properties of a DNA duplex, the vector $\mathbf{u}_i$ is kept aligned with the local backbone direction by a suitable potential described below.

Compared to atomistic simulations of DNA supercoiling~\cite{Mitchell2011,Pyne2021,Burman2025}, this coarse-grained representation sacrifices atomic detail -- particularly the ability to capture torsion-induced melting into single-stranded DNA -- but enables the study of significantly longer chains and timescales.

The translational motion of each bead obeys the Langevin equation:
\begin{equation}\label{eq:langevin}
    m \ddot{\mathbf{r}}_i = -\gamma \dot{\mathbf{r}}_i - \nabla_{\mathbf{r}_i} U + \boldsymbol{\eta}_i(t),
\end{equation}
where $m$ is the bead mass, $\gamma$ the translational friction coefficient, $U$ the total potential energy, and $\boldsymbol{\eta}_i(t)$ a Gaussian stochastic force satisfying the usual fluctuation-dissipation relations: $\langle \boldsymbol{\eta}_i(t) \rangle = 0$ and $\langle \eta_{i,a}(t) \eta_{j,b}(t') \rangle = 2\gamma k_B T \delta_{ij} \delta_{ab} \delta(t-t')$, where $a,b$ label the coordinates in the 3D space.

Each bead also carries an internal \textit{spin angular momentum} $\mathbf{l}_i$, which governs the time evolution of its local frame $\{\mathbf{f}_i, \mathbf{v}_i, \mathbf{u}_i\}$. The dynamics of $\mathbf{l}_i$ follows a rotational Langevin equation:
\begin{align} \label{eq:evol_phi_generic}
    \dot{\mathbf{l}}_i = I_b \dot{\boldsymbol{\omega}}_i &= \boldsymbol{\tau}_{U,i} + \tau_a \, \mathbf{u}_i \, \delta_{i,0} 
    - \gamma_{\rm rot} \boldsymbol{\omega}_i + \boldsymbol{\xi}_i(t),
\end{align}
where $I_b = m \sigma^2 / 10$ is the moment of inertia of a spherical bead, $\boldsymbol{\omega}_i$ is the \textit{spin angular velocity}, and $\boldsymbol{\xi}_i(t)$ is a Gaussian stochastic torque with zero mean and covariance $\langle \xi_{i,a}(t) \xi_{j,b}(t') \rangle = 2 \gamma_{\rm rot} k_B T \delta_{ij} \delta_{ab} \delta(t-t')$. 
The rotational friction is set to $\gamma_{\rm rot} = s \gamma I_b / m$ with $s = 10/3$, ensuring the correct rotational diffusivity for a sphere. 
The deterministic torque $\boldsymbol{\tau}_{U,i}$ derives from the potential $U$. 
An additional external torque of fixed magnitude $\tau_a$ is applied exclusively to the first bead
(i= 0) along its instantaneous axis $\mathbf{u}_0$ to drive twist propagation along the chain. Note
that both chain ends, including the one to which the torque is applied are not clamped but
can move through space.
From now on, we refer to the applied torque as an \textit{active torque}.

The total potential energy is
\begin{equation}
    U = U_{\rm LJ} + U_{\rm FENE} + U_{\rm bend} + U_{\rm twist} + U_{\rm align}.
\end{equation}

Excluded-volume interactions are modeled via a truncated-and-shifted Lennard-Jones potential:
\begin{align}
    U_{\rm LJ} &=\sum_{i<j} \left\{4\left[ \left(\frac{\sigma}{r_{ij}}\right)^{12} 
    - \left(\frac{\sigma}{r_{ij}}\right)^6 \right] 
    + 1\right\} k_B T\theta\!\left(2^{1/6}\sigma - r_{ij}\right), \label{eq:u_lj}
\end{align}
with $r_{ij} = |\mathbf{r}_i - \mathbf{r}_j|$ and thermal energy $k_B T = 4.1$~pN$\cdot$nm, and
where $\theta$ denotes the Heaviside function.

Chain connectivity is enforced through the finitely extensible nonlinear elastic (FENE) potential \cite{Kremer_1990}:
\begin{align}
    U_{\rm FENE} &= -\frac{\kappa_f}{2}  \sum_{i=0}^{N-2} \bigg(\frac{R_0}{\sigma}\bigg)^2
    \ln \left[ 1 - \left(\frac{r_{i+1,i}}{R_0}\right)^2 \right],
    \label{eq:u_fene}
\end{align}
with $\kappa_f = 30 k_B T$ and $R_0 = 1.6\sigma$.

Bending rigidity is introduced via
\begin{align}
    U_{\rm bend} &= \kappa_b \sum_{i=0}^{N-2} (1 - \cos\theta_i), \label{eq:u_bending}
\end{align}
where $\cos\theta_i = \frac{\textbf{t}_{i} \cdot \textbf{t}_{i+1}}{|\textbf{t}_{i}||\textbf{t}_{i+1}|}$ and $\mathbf{t}_i = \mathbf{r}_{i+1} - \mathbf{r}_i$. 
The bending stiffness $\kappa_b = k_B T l_b / \sigma = 20 k_B T$ yields a nominal persistence length $l_b = \kappa_b \sigma / k_B T = 50$~nm.

Torsional stiffness is modeled using Euler angles $\alpha_i$, $\beta_i$, $\gamma_i$ that describe the relative rotation between consecutive local frames~\cite{Chirico1994,Brackley2014}, see Supplementary Figure S1:
\begin{align}
    U_{\rm twist} &= \kappa_t \sum_{i=0}^{N-2} \left[1 - \cos(\alpha_i + \gamma_i)\right], \label{eq:u_twisting}
\end{align}
with $\kappa_t = k_B T l_t / 2\sigma = 30 k_B T$, corresponding to a twist persistence length $l_t = 2\kappa_t \sigma / k_B T = 150$~nm.

Alignment between the bond vector $\mathbf{t}_i$ and the local axis $\mathbf{u}_i$ is enforced by
\begin{align}
    U_{\rm align} &= \kappa_a \sum_{i=0}^{N-2} (1 - \cos\psi_i), \label{eq:u_align}
\end{align}
where $\cos\psi_i = \frac{ (\mathbf{t}_i \times \mathbf{v}_i) \cdot (\mathbf{t}_i \times \mathbf{v}_{i+1}) }
{ |\mathbf{t}_i \times \mathbf{v}_i| \, |\mathbf{t}_i \times \mathbf{v}_{i+1}| }$. 
As previously mentioned, a large alignment constant $\kappa_a = 90 k_B T$ ensures near-perfect collinearity between $\mathbf{u}_i$ and $\mathbf{t}_i$, so that $\theta_i \approx \beta_i$. 
Henceforth, we refer to $\beta_i$ and $\alpha_i + \gamma_i$ as the \textit{bending} and \textit{twist angles}, respectively.

A phase angle $\phi_i$ is defined for each bead as $\phi_i = \sum_{j=0}^{i} (\alpha_j + \gamma_j)$, so that $\alpha_i + \gamma_i = \phi_{i+1} - \phi_i$. 
The total twist stored in the chain is then ${\rm Tw} = \frac{1}{2\pi} \sum_{i=0}^{N-2} (\alpha_i + \gamma_i)$.

The torque derived from the potential is~\cite{Brackley2014}
\begin{equation}
    \boldsymbol{\tau}_{U,i} = -\kappa_a \, \mathbf{u}_i \times \mathbf{t}_i 
    - \kappa_t (\mathbf{H}_i - \mathbf{H}_{i-1}),
\end{equation}
with $\mathbf{H}_i = \frac{ \mathbf{u}_{i+1}\times\mathbf{u}_i \cos(\alpha_i + \gamma_i) 
- \mathbf{f}_{i+1}\times\mathbf{f}_i - \mathbf{v}_{i+1}\times\mathbf{v}_i }{1 + \mathbf{u}_{i+1}\cdot\mathbf{u}_i}$.

Note that in this model, the typical parameters to tune in order to reproduce the DNA properties (e.g., in specific ionic conditions) are the elasticity parameters, $l_b$ and $l_t$, and the effective filament thickness, $\sigma$. Here, such values are chosen to match DNA in physiological conditions, $\sigma=2.5$nm,  $l_b=50nm,$ and $l_t=150nm$.
All quantities are expressed in units of $m$, $\sigma$, and $k_B T$. The molecular dynamics time unit is $t_{\rm MD} = \sigma \sqrt{m / k_B T}$, and a timestep of $0.01\, t_{\rm MD}$ is used. As customary, the friction coefficient is set to $\gamma = 2$ \cite{Kremer_1990}, yielding $\gamma_{\rm rot} = 0.66$.
Equations of motion are integrated using the velocity-Verlet algorithm in \textsc{LAMMPS}~\cite{LAMMPS}, with custom force terms implemented following Ref.~\cite{Brackley2014}.

We simulate DNA chains of 2, 5, and 10~kbp, corresponding to $N = 264$, 679, and 1378 beads, and to contour lengths of $0.66,\, 1.70,\, 3.4\, \mu m$, respectively. Initial configurations are generated via Monte Carlo annealing to obtain equilibrated, untwisted backbones of the duplex. 
The active torque is varied in the range $\tau_a \in [0, 10]\, k_B T$ (i.e., $[0, 41]$~pN$\cdot$nm), and is applied along $\mathbf{u}_0$, the chain axis at the first bead. For each chain length and torque, we considered 100 different initial configurations.

A nominal mapping from simulation time to real physical units can be performed as follows. The relevant microscopic characteristic timescale is given by the nominal self-diffusion time of a single monomer in the chain.
Equivalently, one can define the mapping by matching the self-diffusion coefficient of an isolated bead $D=\frac{k_BT}{\gamma}=0.5\frac{\sigma^2}{t_{MD}}$ with  the one estimated by applying Stokes' law to a bead of diameter $\sigma=2.5 nm$ immersed in a fluid of viscosity equal to water, 1cP. Setting $T = 300 K$, the matching, which neglects hydrodynamic interactions, yields  $t_{MD} \approx 40$ ns. Thus, our simulations, which last typically  $10^8t_{MD}$, correspond to $4$s.

\section{Perfectly straight chain limit} \label{linear_polymer}

As a preliminary step, we consider the simplified case of a perfectly straight chain: this idealized limit could be achieved, for instance, by imposing an infinitely large bending rigidity. In this case, all bond vectors \(\mathbf{t}_i\) and local axes \(\mathbf{u}_i\) remain strictly parallel, and the spin angular velocity \(\boldsymbol{\omega}_i\) is collinear with \(\mathbf{u}_i\). Consequently, the phase angle \(\phi_i\) evolves as \(\dot{\phi}_i = {\omega}_i\).
Under these conditions, the rotational equation of motion, Eq.~\eqref{eq:evol_phi_generic}, simplifies significantly. 
Averaging over stochastic noise, we obtain:

\begin{equation} \label{eq:phi_evolution_cyl}
\begin{aligned}
    I_b \ddot{\phi}_0 &= \kappa_t \sin(\phi_1 - \phi_0) + \tau_a - \gamma_{\rm rot} \dot{\phi}_0, \\
    I_b \ddot{\phi}_i &= \kappa_t \left[ \sin(\phi_{i+1} - \phi_i) - \sin(\phi_i - \phi_{i-1}) \right] 
                     - \gamma_{\rm rot} \dot{\phi}_i, \quad (1 \leq i \leq N-2), \\
    I_b \ddot{\phi}_{N-1} &= -\kappa_t \sin(\phi_{N-1} - \phi_{N-2}) - \gamma_{\rm rot} \dot{\phi}_{N-1},
\end{aligned}
\end{equation} 
where \(\phi_{i+1} - \phi_i = \alpha_i + \gamma_i\) is the local twist angle.

At steady state, all beads rotate with the same angular velocity, \(\dot{\phi}_i = \omega_L\). Substituting into Eq.~\eqref{eq:phi_evolution_cyl} and summing over all beads yields:
\begin{equation} \label{omega}
    \omega_L = \frac{\tau_a}{N \gamma_{\rm rot}}.
\end{equation}
The steady-state twist profile is then
\begin{equation} \label{eq:twist_distribution_cyl}
    \sin(\phi_{i} - \phi_{i-1}) = -\frac{\tau_a}{\kappa_t} \left(1 - \frac{i}{N}\right).
\end{equation}
Thus, the stationary angular velocity increases linearly with the applied torque and decreases inversely with chain length. For positive $\tau_a$, a negative twist accumulates along the filament, with magnitude proportional to \(\tau_a / \kappa_t\). Note that the twist profile decays linearly from the driven end (\(i=0\)) to the free end (\(i=N-1\))  for $| \tau_a/\kappa_t| \le 1$ (e.g. in a small-torque regime), where  $\sin(\phi_{i+1} - \phi_i)\approx \phi_{i+1} - \phi_i$.

We can also take the continuum limit of Eqs.~\eqref{eq:phi_evolution_cyl}, which is
\begin{equation} \label{eq:phi_evolution_cylinder}
    \hat{I}_b \partial_t^2 \phi(s,t) = \frac{k_B T l_t}{2} \partial_s^2 \phi(s,t) 
    - \hat{\gamma}_{\rm rot} \partial_t \phi(s,t) + \tau_a \delta(s),
\end{equation}
with \(s = i\sigma\)  the arc-length coordinate,  \(\hat{I}_b = I_b / \sigma\) and \(\hat{\gamma}_{\rm rot} = \gamma_{\rm rot} / \sigma\) the inertia and rotational friction per unit length, respectively.

Defining the local twist density as
\begin{equation}
    \rho_{\rm tw}(s,t) = \frac{1}{2\pi} \partial_s \phi(s,t),
\end{equation}
and neglecting inertial effects—which are negligible for DNA dynamics in solution due to the extremely low rotational Reynolds number~\cite{Phillips2012}, we obtain the diffusion equation for twist density:
\begin{equation} \label{eq:twist_evolution_continuous_cyl}
    \partial_t \rho_{\rm tw}(s,t) = D_{\rm tw} \partial_s^2 \rho_{\rm tw}(s,t) 
    + \frac{\tau_a}{2\pi \hat{\gamma}_{\rm rot}} \delta'(s),
\end{equation}
with twist diffusivity \(D_{\rm tw} = k_B T l_t / (2 \hat{\gamma}_{\rm rot})\), and $\delta'(s)$ being the derivative of the Dirac delta function. Note that for a finite polymer with boundaries $s\in [0,N\sigma]$, one needs to set zero accumulation of twist  outside these boundaries, $\rho_{\rm tw}=0$. Equivalently, one can recast Eqs.~\ref{eq:phi_evolution_cylinder}  and ~\ref{eq:twist_evolution_continuous_cyl} to their homogeneous forms (i.e.~without involving the delta function) supplemented with the boundary conditions $\rho_{\rm tw}(0,t)=-\frac{\tau_a}{2\pi D_{\rm tw}\hat{\gamma}{\rm rot}}$, and $\rho_{\rm tw}(N\sigma,t)=0$.

In steady state, the solution recovers the linear twist gradient of Eq.~\eqref{eq:twist_distribution_cyl} in the small-angle limit, \(\sin(\phi_{i+1}-\phi_i) \approx \phi_{i+1}-\phi_i\).

The perfectly straight chain thus serves as a minimal reference model in which torsional stress propagates exclusively via \textit{twist diffusion}, resulting in uniform rigid-body rotation and a stationary linear gradient of twist.

\section{The three-dimensional polymer case} \label{results}
Having established the behavior in the limit of a straight chain, we now consider the case in which the DNA chain can adopt a fully three-dimensional configuration due to the combination of torque application and thermal fluctuations (Fig.~\ref{fig:configurations}). 
This unlocks additional relaxation channels -- local bending and plectoneme formation -- that dramatically reshape the propagation and dissipation of torsional stress along the filament.
Hereafter, we systematically investigate the structural and dynamical response of the polymer by varying both the applied active torque magnitude and the DNA length.

\textbf{Configurations at different torques}. Fig.~\ref{fig:configurations}(b) shows configurations of polymers of equal length ($2 {\rm kbp}$) under different torque values $\tau_a$ (low to high values from bottom to top rows) and at various times (different columns, time increasing from left to right), see also Supplementary Movies S1-S5. We start from the same initially equilibrated configuration in the absence of torque, and, starting from time $t=0$, apply the active torque at one end. As time progresses, we find two markedly different behaviors. For $\tau_a < \tau_{th}$, with $\tau_{th}\sim 2$ an active torque threshold, the chain remains in a swollen phase, with its    morphology  essentially unaffected by the imposed torsional stress. Instead, for $\tau_a \ge \tau_{th}$ the stress drives the formation of a plectoneme near the driven end, which gradually expands over time. Eventually, other plectonemes can form further from the driven end. Typically, plectonemes become more compact with increasing active torque, and at long times, we reach a steady state with one (or, less frequently, more than one) persisting plectoneme (see e.g., $t=1 \cdot 10^{6}t_{MD}$).

Thus, in three dimensions, the chain has two channels through which the continuously introduced torsional stress can be dissipated. First, the chain can accumulate twist, which may be released through the axial motion of all its beads, because the end opposite to the one where torque is applied is free. 
Second, when $\tau_a$ exceeds a critical threshold, part of the excess torsional stress is released through local bending, giving rise to writhe. 
Regions of high curvature act then as nucleation sites for the formation of plectonemes. This phenomenon was also previously described in~\cite{Wada2009}, where the complementary fixed angular velocity ensemble (rather than fixed torque) was considered.

\begin{figure}
    \centering
    \includegraphics[width=0.75\linewidth]{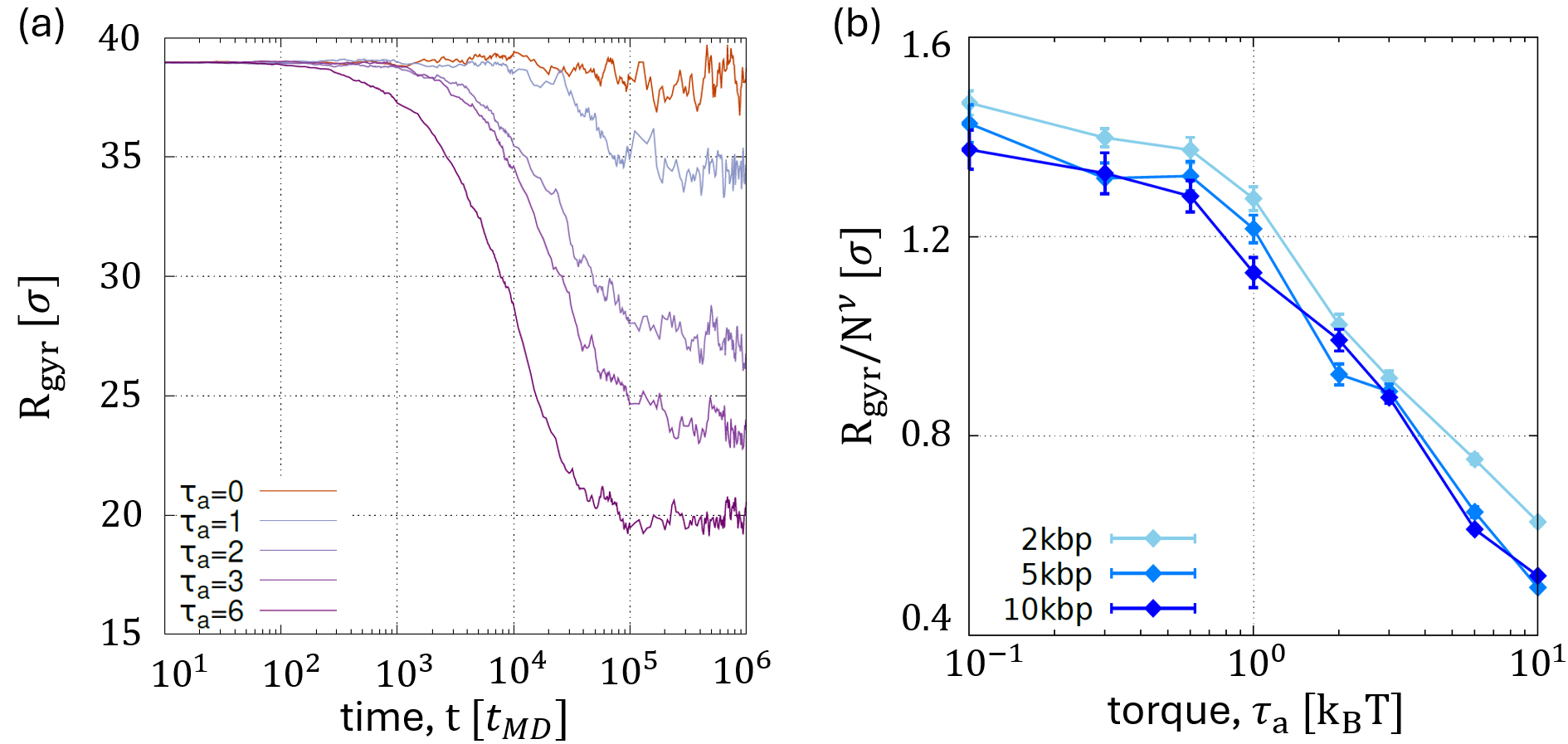}
    \caption{(a) Radius of gyration as a function of time, for different torques $\tau_a=0,1,2,3,6$, and fixed chain length $N=2\rm kbp$.(b) Rescaled gyration radius, $R_{gyr}/N^\nu$, for various chain lengths and as a function of the active torque $\tau_a$. The Flory scaling exponent was set to $\nu=0.588$.}
    \label{fig:rgyr_ete}
\end{figure}

\textbf{Polymer morphology}.
To characterize the overall morphology of the polymer chain, we consider the \textit{radius of gyration} $R_{gyr} =\sqrt{\sum_{i=1}^{N} (\mathbf{r}_i - \mathbf{r}_{\text{cm}})^2/N}$. Fig. \ref{fig:rgyr_ete}(a) shows the time evolution of $R_{gyr}$ for various active torques at $N=2$kbp: all curves reach a stationary regime at long times $t>10^5 t_{MD}$,  for which they become constant. The value of $R_{gyr}$ in this regime decreases with increasing $\tau_a$.

Fig. \ref{fig:rgyr_ete}(b) shows values of $R_{gyr}$ in the stationary regime for all chain lengths as a function of $\tau_a$.
Independent of the chain length, we observe that $R_{gyr}$ remains nearly constant in the swollen phase, $\tau_{th} < 2$. After this threshold, $R_{gyr}$ decreases with increasing $\tau_a$, due to the formation of plectonemes which compact the chain. Notably, curves collapse well for all torque values when  scaling $R_{gyr}$ by $N^{\nu}$, with $\nu \simeq 0.588$ being the Flory exponent for a self-avoiding random walk. Minor differences, found mainly at low torques, can be attributed to finite-size effects\cite{Besold2000}.

A simple scaling argument to predict the scaling of the threshold torque $\tau_{th}$ with the system parameters in our force-free setup can be built by comparing the energetic costs of twisting and bending, in a similar way to what is done to study the scaling of the buckling transition with constant force and supercoiling density. First, the cost of twisting a segment of length $l$ can be estimated as $E_{twist}\sim \frac{\kappa_tl(\alpha_i+\gamma_i)^2}{2\sigma}  \sim \frac{\tau_a^2 l}{8\sigma\kappa_t}$, where we have estimated an average  twist angle of $\alpha_i+\gamma_i\sim \frac{\tau_a}{2\kappa_t}$ from Eq.~\ref{eq:twist_distribution_cyl}. Second, the cost of creating a plectoneme (or loop) of size $l$ is instead $E_{bend}\sim\frac{\sigma\kappa_b}{2l}$. The transition $\tau_{th}$ between bending being more favorable than twisting (costing less free energy, $\tau_a>\tau_{th}$) and vice versa occurs at $\tau_{th}\sim \sqrt{\frac{4\sigma^2\kappa_t \kappa_b}{l^2}}$.
If we assume, as appears reasonable, that $l\sim l_b$, we then obtain $\tau_{th}\sim 2k_BT \sqrt{\frac{\kappa_t}{\kappa_b}}\sim 2 k_BT$, which is of the same order as the one found numerically.

\textbf{Spin angular velocity and global chain rotation}. 
We now examine the profile of the average spin angular velocity projected along the local chain direction, $\boldsymbol{\omega}_i \cdot \mathbf{u}_i$, computed in the stationary regime, to assess how effectively the spin imparted at the driven end ($i=0$) propagates along a three-dimensional polymer chain. In other words, it quantifies the extent to which the chain is rotating (spinning) about its own local tangent axis—i.e., torsional propagation along the backbone—see also Supplementary Figure S1.

\begin{figure}
    \centering
    \includegraphics[width=0.75\linewidth]{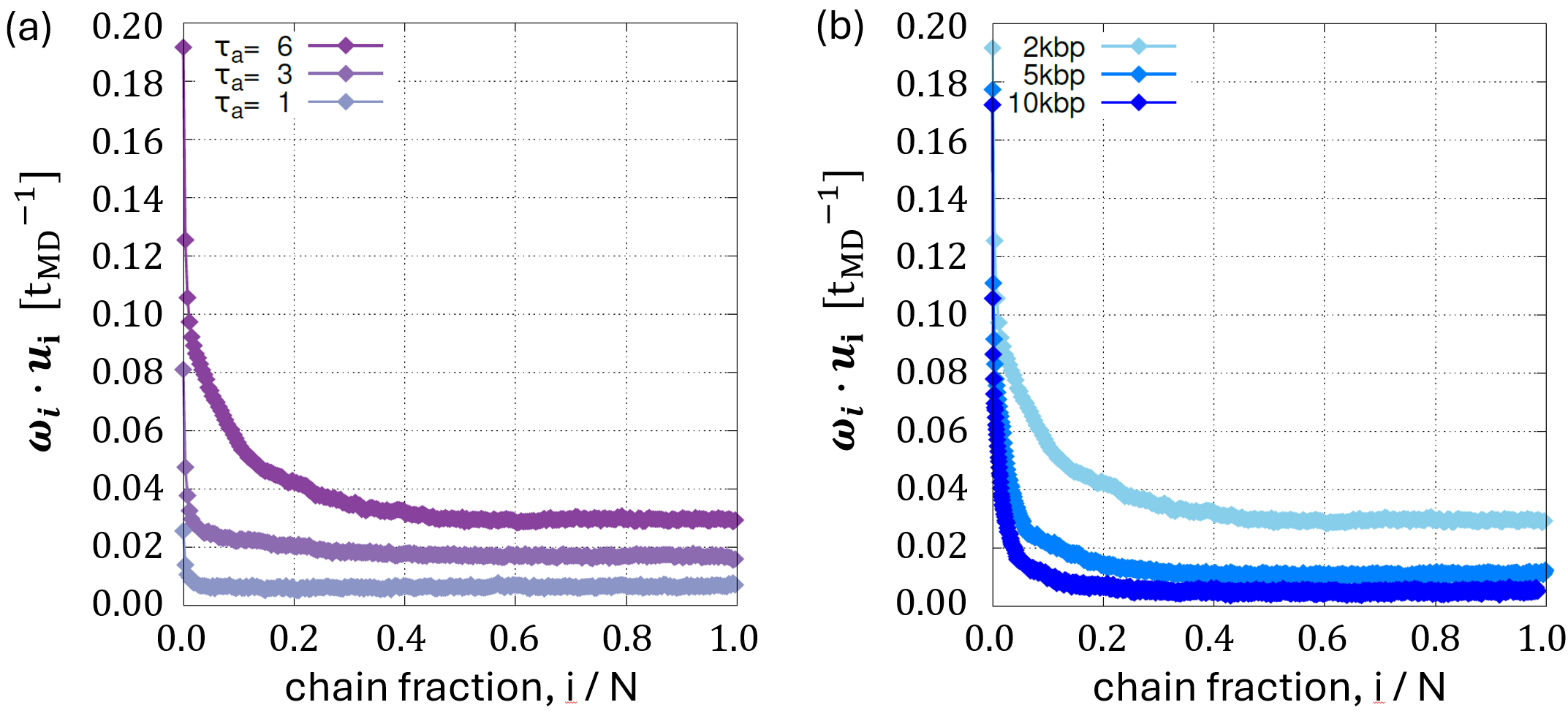}
    \includegraphics[width=0.75\linewidth]{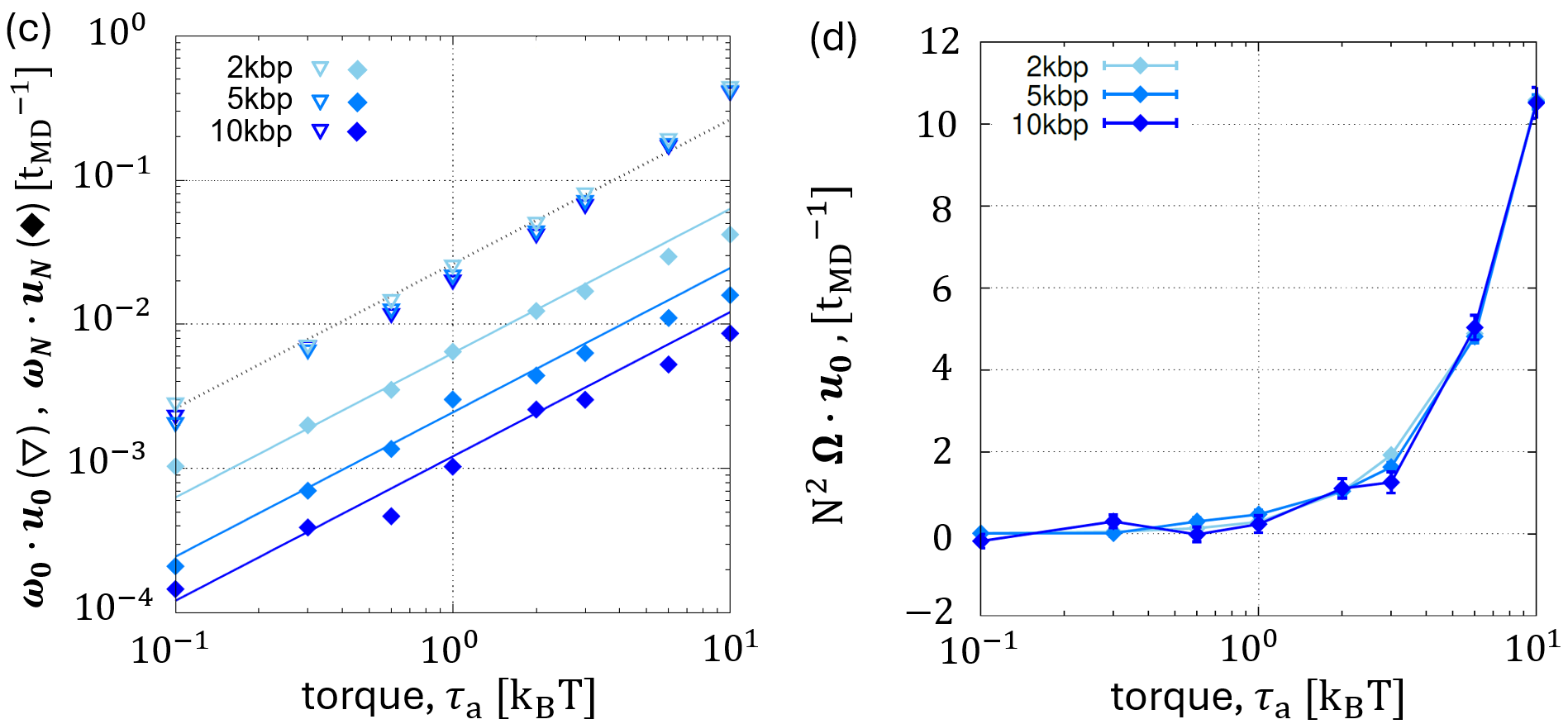}
    \caption{(a) Spin angular velocity profiles in the direction tangential to the chain, $\boldsymbol{\omega}_i \cdot \mathbf{u}_i$ as a function of the chain fraction $i/N$, for different torques $\tau_a=1,3,6$ and at fixed chain length $2 \mathrm{kbp}$.
    (b) $\boldsymbol{\omega}_i \cdot \mathbf{u}_i$ as a function of $i/N$ for different chain length at fixed torque $\tau_a=6$.
    (c) $\boldsymbol{\omega}_0 \cdot \mathbf{u}_0$ (triangles), and $\boldsymbol{\omega}_N \cdot \mathbf{u}_N$, diamonds, as a function of the applied torque for various chain lengths. The solid lines are the theoretical values $\omega_L=\frac{\tau_a}{N\gamma_{rot}}$, Eq.~\ref{omega}, computed for the different chain lengths. The dotted black line, obtained from a fit, is $2.3 \cdot 10^{-2} \tau_a$.
    (d) Orbital angular velocity $\boldsymbol{\Omega}$ of the chain projected on the direction of the applied torque $\mathbf{u}_0$, as a function of the applied torque for various chain lengths. $\boldsymbol{\Omega}$ has been scaled by the chain length squared $N^2$.}
    \label{fig:axial_velocity}
\end{figure}

Fig.~\ref{fig:axial_velocity}(a) shows this profile for the shortest chain ($N = 2 \, \rm{kbp}$) and three different values of torque $\tau_a$. For any $\tau_a$, the spin velocity from the point of torque application ($i=0$) towards a finite plateau value at the free end ($i=N$). This behavior is consistent with previous observations in Ref.~\cite{Wada2009}. 
We observe a systematic increase in the overall spin velocity with increasing $\tau_a$. Additionally, the decay becomes progressively slower at higher torque.

In contrast, fixing $\tau_a$ and varying the chain length [Fig.~\ref{fig:axial_velocity}(b)] reveals that the spin velocity at the anchored end, $\boldsymbol{\omega}_0 \cdot \mathbf{u}_0$, is independent of $N$, while the plateau value at the free end, $\boldsymbol{\omega}_N \cdot \mathbf{u}_N$, decreases with chain length.

Fig.~\ref{fig:axial_velocity}(c) reports the dependence of both boundary values on $\tau_a$. We find linear scaling in both cases
\begin{equation}
    \boldsymbol{\omega}_0 \cdot \mathbf{u}_0 \sim   2.3 \cdot 10^{-2} \tau_a, \quad
    \boldsymbol{\omega}_N \cdot \mathbf{u}_N \sim \omega_L,
    \label{eq:spin_scaling}
\end{equation}
represented by the dotted line and the solid lines, respectively, where $\omega_L$ is the predicted spin angular velocity for a straight chain given by Eq.~\ref{omega}. 
Interestingly, in Ref.~\cite{Wada2009} the presence of hydrodynamics affects differently the scaling of $\boldsymbol{\omega}_N \cdot \mathbf{u}_N$, which remains equal to $\boldsymbol{\omega}_0 \cdot \mathbf{u}_0$ in the swollen phase, and becomes proportional to $N^{-2/3}$ in the plectonemic phase.

Beyond local spinning about its contour, the active torque also induces a global orbital rotation of the entire molecule, where the entire coil whirls as a single unit around a rotational axis, see Supplementary Figure S1. To characterize this collective motion, we computed the orbital angular velocity $\boldsymbol{\Omega}$ with respect to the polymer center of mass, through the following formula:
\begin{equation}
    \mathbf{L} = \mathbf{I} \cdot \boldsymbol{\Omega},
    \label{eq:orbital_momentum}
\end{equation}
where $\mathbf{I}$ is the moment-of-inertia tensor and $\boldsymbol{L}$ is the global angular momentum. Note that we obtain $\boldsymbol \Omega$ by computing directly both $\mathbf{I}$ and $\boldsymbol{L}$ in a given conformation, and inverting the tensor relation.

Fig.~\ref{fig:axial_velocity}(d) shows the component of $\boldsymbol{\Omega}$ projected along $\mathbf{u}_0$, the direction along which the global rotation occurs, and averaged at steady state over different conformations. Global orbital rotation is apparent for $\tau_a \ge \tau_{th}$, with all curves collapsing onto a master curve when rescaling $\boldsymbol{\Omega}$ by $N^2$, implying that $\boldsymbol{\Omega}\propto N^{-2}$ at fixed $\tau_a$.

\textbf{Twist and bend accumulation}.
We now turn to analyze how twist and bend accumulate along the chain, first over time (globally), and then along the chain (locally). While twist describes the accumulation of internal torsion along the chain, bend provides a first intuitive proxy to evaluate how part of the accumulated twist converts into writhe.

\begin{figure}
    \centering
    \includegraphics[width=0.75\linewidth]{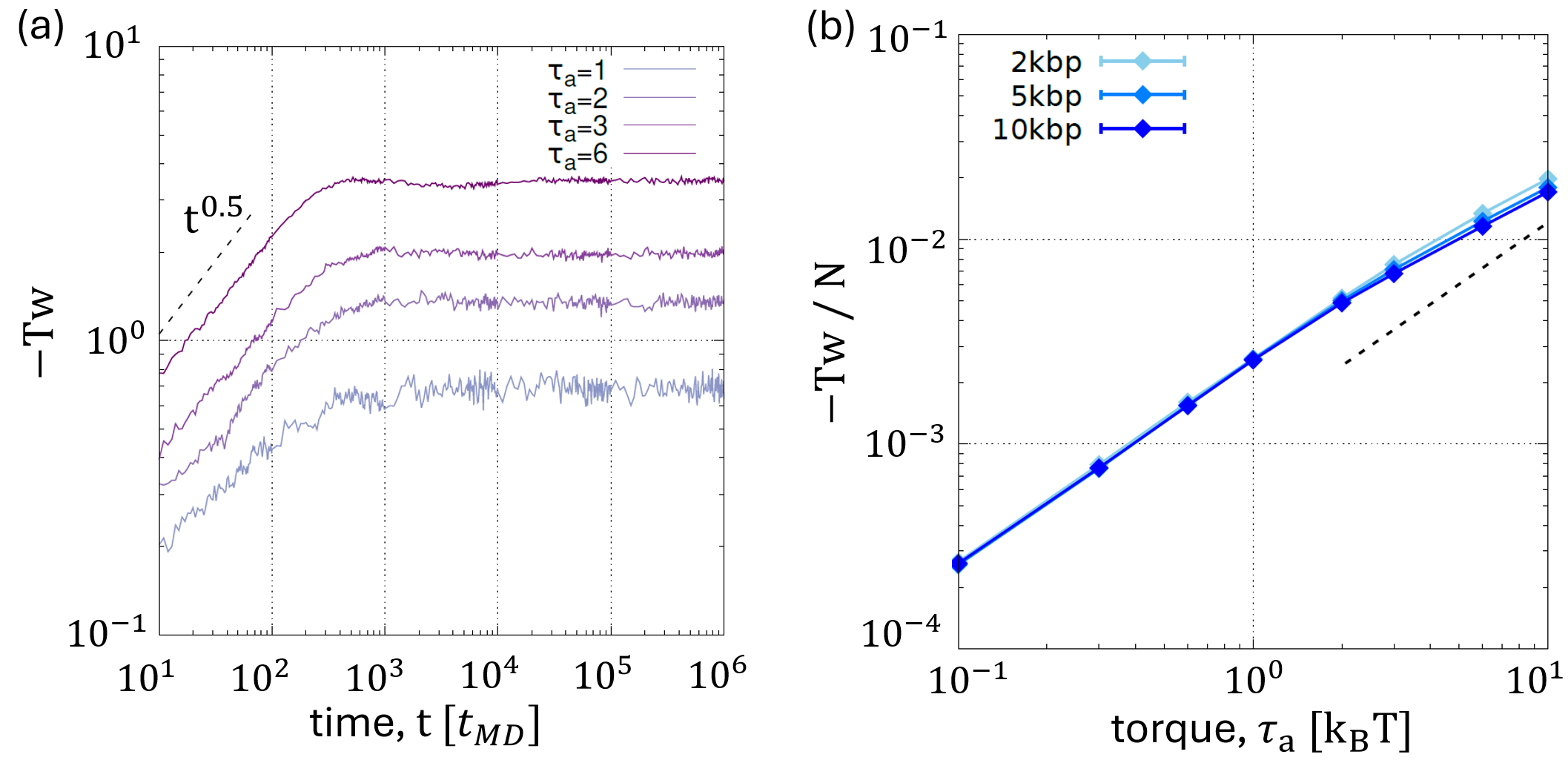}
    \includegraphics[width=0.75\linewidth]{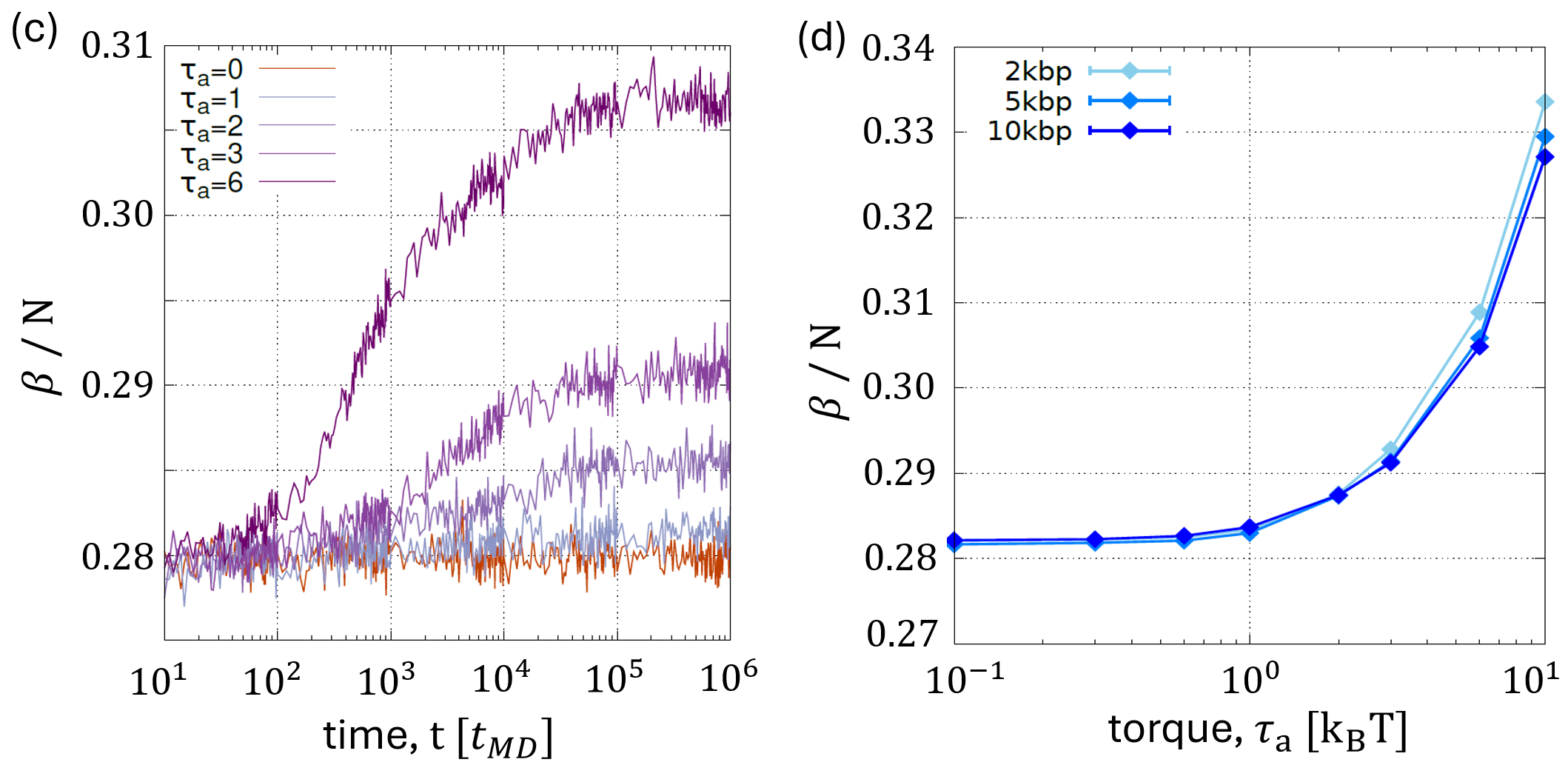}   
    \caption{(a) Total twist ${\rm Tw}$ as a function of time for various active torques $\tau_a$. The dashed line indicates the initial diffusive scaling ${\rm Tw} \propto t^{0.5}$. ${\rm Tw}$ is negative for positive applied torques. (b) Total twist accumulated along the chain, rescaled over the chain length $N$, as a function of $\tau_a$ and for various $N$. The dashed line is a linear trend in $\tau_a$. (c) Total bending angle $\beta$ as a function of time for the same $\tau_a$ values as in (a). (d) $\beta$, rescaled over $N$, as function of $\tau_a$ for various $N$.}
    \label{fig:twist_along_chain}
\end{figure}

The global accumulation of twist can be computed as the sum over twist angles
\begin{equation}
    {\rm Tw}=\frac{1}{2\pi}\sum_i (\alpha_i+\gamma_i), 
    \label{totalTw}
\end{equation}
while the total bend can be estimated as the sum over all bending angles
\begin{equation}
    \beta=\sum_i \beta_i.
\end{equation}
Fig. \ref{fig:twist_along_chain}(a)-(c) shows the global accumulation of twist and bend over time, for $N=$2 kbp and several $\tau_a$. Note that we plot $-{\rm Tw}$, as a positive active torque yields a negative accumulation of twist, as found for the straight chain. We also found that in the initial regime $-{\rm Tw}\propto t^{0.5}$, compatible with a diffusive behavior for the twist propagation, which is expected for a straight chain from  Eq.~\eqref{eq:twist_evolution_continuous_cyl}. For both ${\rm Tw}$ and $\beta$ curves reach saturation, but for twist it occurs much faster ($t\sim 10^3t_{MD}$) than for bending ($t\sim 10^5t_{MD}$); the latter timescale is compatible with the saturation time of $R_{gyr}$. Using the nominal time-to-physical-unit mapping, the indicative characteristic saturation times of twist and bending lie in the range $0.04-4$ ms, in agreement with Ref.~\cite{wan2022two}.
As the dynamics of bending depends on plectoneme nucleation events~\cite{Daniels2011}, we do not expect a specific exponent to govern its growth law over time. 

Fig. \ref{fig:twist_along_chain}(b)-(d) show the stationary values of ${\rm Tw}$ and $\beta$ as a function of $\tau_a$ for different $N$. We found that both quantities are proportional to $N$ at all $\tau_a$, meaning that the longer the chain is, the higher the amount of torsion and bending accumulated. Additionally, 
the global twist increases as ${\rm Tw}\propto \tau_a$ up to $\tau_a\sim \tau_{th}$, as expected from Eq.\ref{eq:twist_distribution_cyl}. 
For higher ${\tau_a}$, instead, we find ${\rm Tw}$ starts to increase sublinearly, implying torque accumulation is hindered by the formation of plectonemes. The average bending angle $\beta/N$ increases with $\tau_a$ when plectonemes are nucleated: curves for different $N$ collapse, underscoring that the behaviour is independent of chain length.

\begin{figure}
    \centering
    \includegraphics[width=0.75\linewidth]{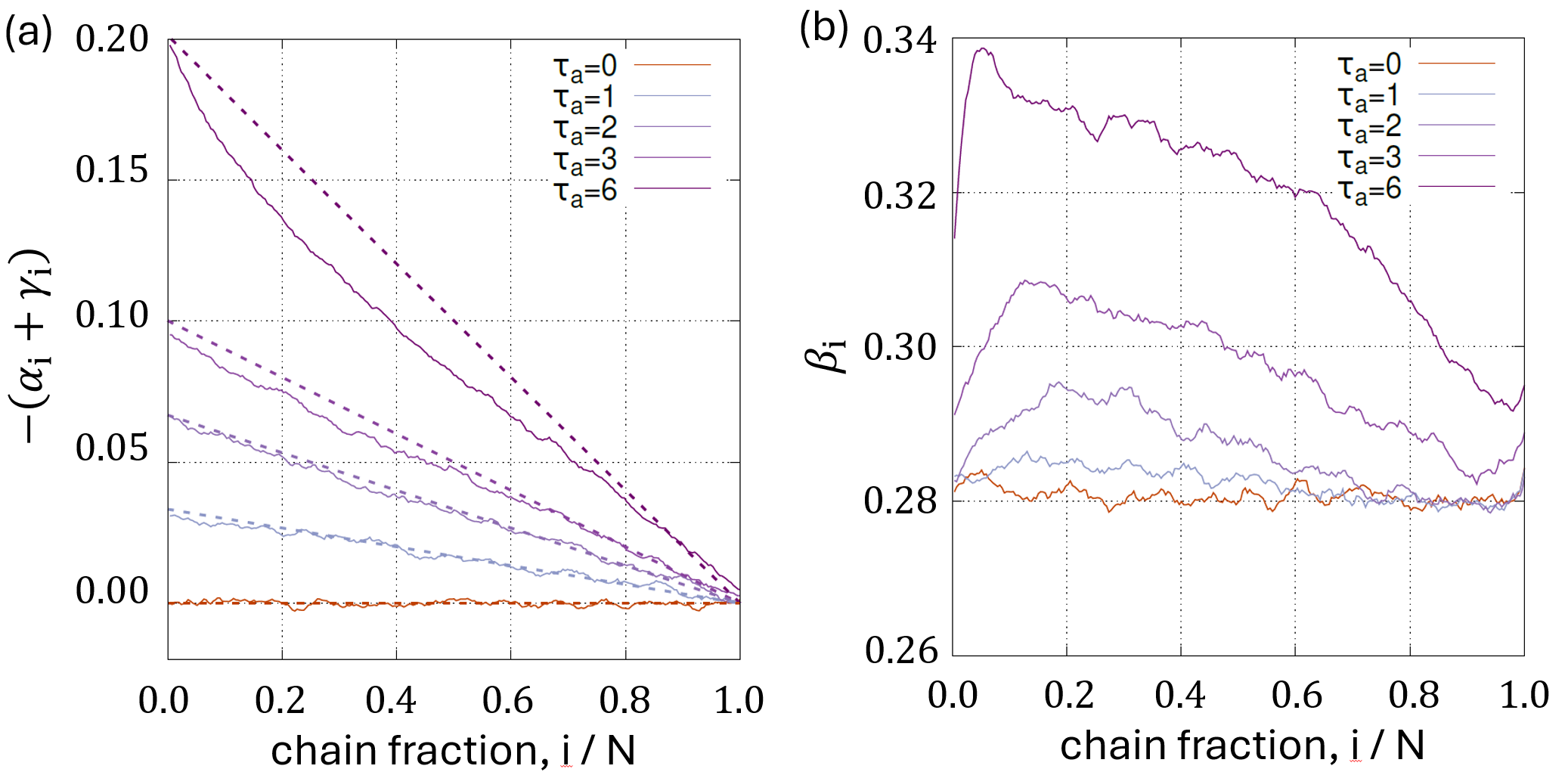}
    \caption{(a) Local twist angle $\alpha_i+\gamma_i$ and (b) local bending angle $\beta_i$ as a function of chain fraction $i/N$,  at different active torques and fixed chain length $2{\rm kbp}$. The   dashed lines in (a) represent, for each active torque value, the theoretical prediction for a straight chain, Eq. \eqref{eq:twist_distribution_cyl}.}
    \label{fig:twist_writhe_along_chain}
\end{figure}

A better insight into twist and bend accumulation can be gained by looking at the twist and bending angles profiles, $\alpha_i+\gamma_i$ and $\beta_i$ for $N=2$kbp, at different torques, Fig.\ref{fig:twist_writhe_along_chain}. 

Regarding twist accumulation, we find that for low active torque the profiles decay linearly to zero with the chain fraction $i/N$ (except for $\tau_a=0$, where it is always zero), which is again expected from Eq.\ref{eq:twist_distribution_cyl}. Instead, at high enough torque, the profile develops a non-linear elbow close to the linear end, which deviates from the straight chain prediction (see comparison with dashed lines, corresponding to Eq.\ref{eq:twist_distribution_cyl} of a straight chain). 

The bend profile starts again at a constant value at $\tau_a=0$ and develops with increasing $\tau_a$ to a maximum near the active torque application point $i=0$. As the torque increases, the peak becomes more prominent and shifts towards $i=0$. In a plectonemic conformation, which is inherently more rigid than the swollen state, the bending angle exceeds that arising from thermal fluctuations alone due to the approximately helical local structure of the plectoneme; in fact,  although the overall plectoneme may appear relatively straight along its centerline, each of the two intertwined duplexes exhibits substantial curvature around this centerline, as each can be locally modeled as a helix with a radius of curvature determined by the pitch and projected radius of the plectoneme (see Supplementary Figure S2).

From a physical perspective, these profiles reflect how the chain stores torsional stress: when the applied torque is small, the chain has sufficient elastic capacity to dissipate stress through local bead rotations, and bending is low. However, once the torque exceeds this limit, the system can no longer accommodate the imposed twist through simple rotation. Instead, local bend develops at sites of high curvature, which act as nucleation points for plectoneme formation -- providing a more efficient pathway to relieve torsional stress. This transition from uniform twist diffusion to localized bending is precisely the mechanism underlying the structural rearrangements observed in Fig. \ref{fig:configurations}, and discussed in \cite{Wada2009}. 

These observations are also connected to the profile shown in Fig.~\ref{fig:axial_velocity}: the drop in spin angular velocity observed at $i=0$ correlates with the deviation of the twist from a linear profile, as well as the rise in bend, i.e., regions of high curvature such as the apex of a plectoneme.
We can thus infer that these bent regions act as a torque-splitting junction, distributing energy into two distinct dynamical channels: internal spin propagation and collective rotation.

\textbf{Plectoneme formation and writhe}. 
We now characterize the plectonemes, structures forming for $\tau_a>\tau_{th}$, by measuring 
the writhe, i.e., the number of self-crossings formed by the polymer. For a  discrete polymer chain, the local writhe for a bead $i$ can be defined as~\cite{Fosado2021}
\begin{equation} \label{eq:writhe_def}
{\rm Wr}_i = \frac{2}{4\pi} \sum_{k=i-l_w}^{i}\sum_{l=i}^{i+l_w}
\frac{(\mathbf{t}_{k,k+1}\times \mathbf{t}_{l,l+1}) \cdot (\mathbf{r}_k-\mathbf{r}_l)}{|\mathbf{r}_k-\mathbf{r}_l|^3},
\end{equation}
which, for each bead $i$, evaluates the $l_w$ preceding and $l_w$ following beads and counts the number of crossings within this window. In our analysis, we fixed $l_w=150$. 
To identify plectonemic regions, while avoiding noise, we first identify separate regions where $|{\rm Wr_i}|>1$, and then extend them to smaller and larger values of $i$ until $|{\rm Wr_i}|$ falls below the threshold value $0.5$, thus defining the two boundaries. 
We can also define the total writhe as:  \cite{Chirico1994}
\begin{align}\label{eq:total_writhe}
    {\rm Wr}=\frac{1}{4\pi} \sum_{k=0}^{N-2} \sum_{l=0}^{N-2} \frac{\left( \mathbf{t}_{k,k+1} \times \mathbf{t}_{l,l+1} \right) \cdot \left( \mathbf{r}_k - \mathbf{r}_l \right)}{|\mathbf{r}_k - \mathbf{r}_l|^3}.
\end{align}

\begin{figure*}
    \centering
    \includegraphics[width=0.95\linewidth]{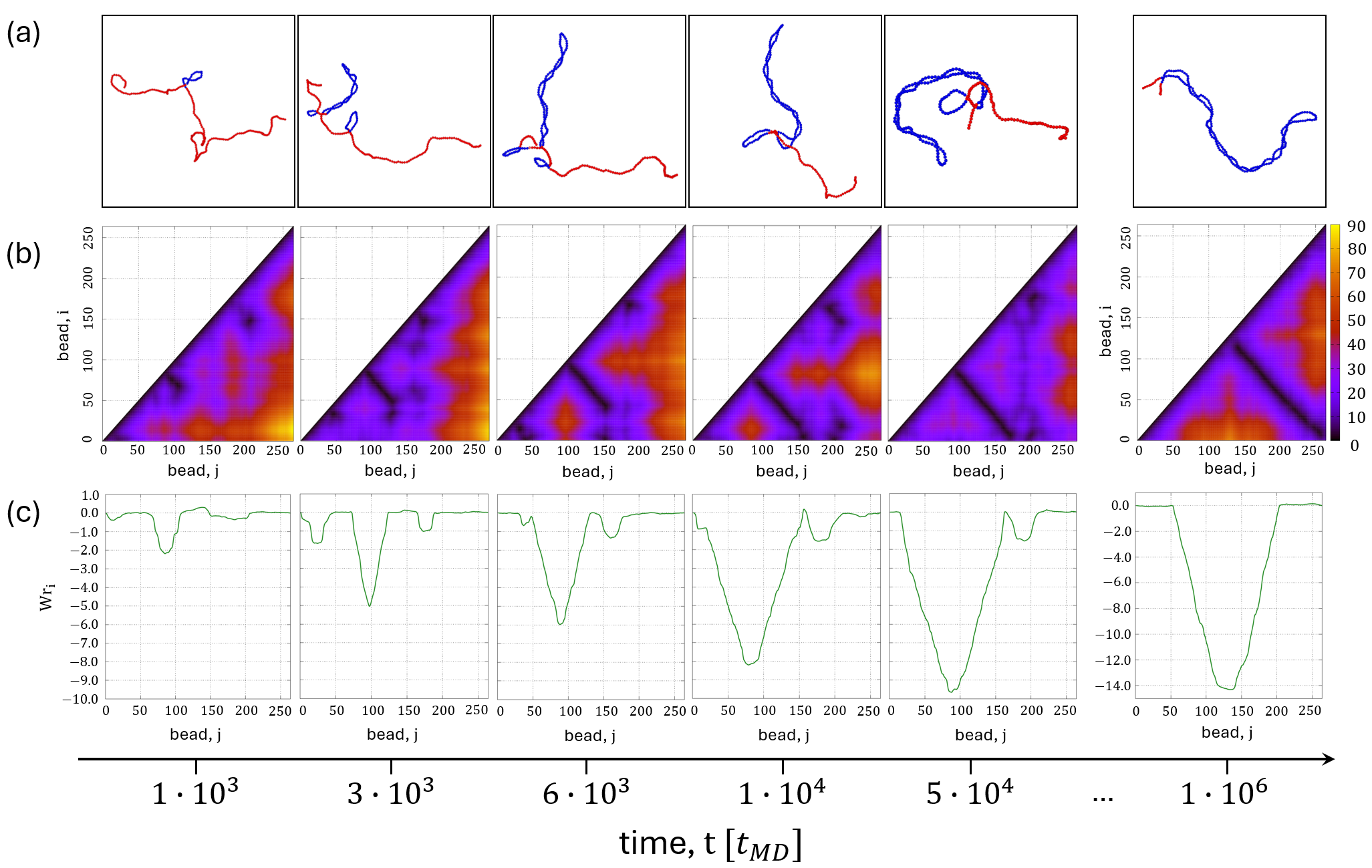}
    \caption{(a) Same configurations as in the top row of Fig.~\ref{fig:configurations}. Regions identified as plectonemes are colored in blue, while the rest of the chain is in red. Time increases from left to right. (b) Corresponding contact maps, and (c) local writhe ${\rm Wr}_i$ as a function of the bead index $i$,  for the same configurations as (a).  $\tau_a=6$ and the chain length is $2 {\rm kbp}$. On the right of (b) is the color bar for the contact map.}
    \label{fig:writhe_contact_map}
\end{figure*}

Fig. \ref{fig:writhe_contact_map}(a) shows the same conformations at different times of Fig.~\ref{fig:configurations}(b), at $\tau_a=6$, now with beads colored in blue in the plectonemic regions, and red otherwise, using the aforementioned detection criteria based on the writhe threshold. 
Fig. \ref{fig:writhe_contact_map}(b) shows the contact map, computed by quantifying the spatial proximity of all pairs of monomers along the polymer chain. A contact between monomers $i$ and $j$ is defined whenever their mutual distance $d_{ij} = \sqrt{(\textbf{r}_i - \textbf{r}_j)^2}$ falls below a chosen cutoff value $d_{cut}=6\sigma$.  
The resulting two-dimensional matrix provides a global view of the three-dimensional organization of the polymer: diagonal elements represent local backbone connectivity. At the same time, off-diagonal clusters of contacts reflect plectonemic structures.
Fig.\ref{fig:writhe_contact_map}(c) shows the local writhe as a function of the chain index, respectively, where we identify plectoneme regions here as those where  ${\rm Wr_i}\le -0.5$. Note that a positive $\tau_a$ generates a writhe which is negative in value.

\begin{figure}
    \centering
    \includegraphics[width=0.6\linewidth]{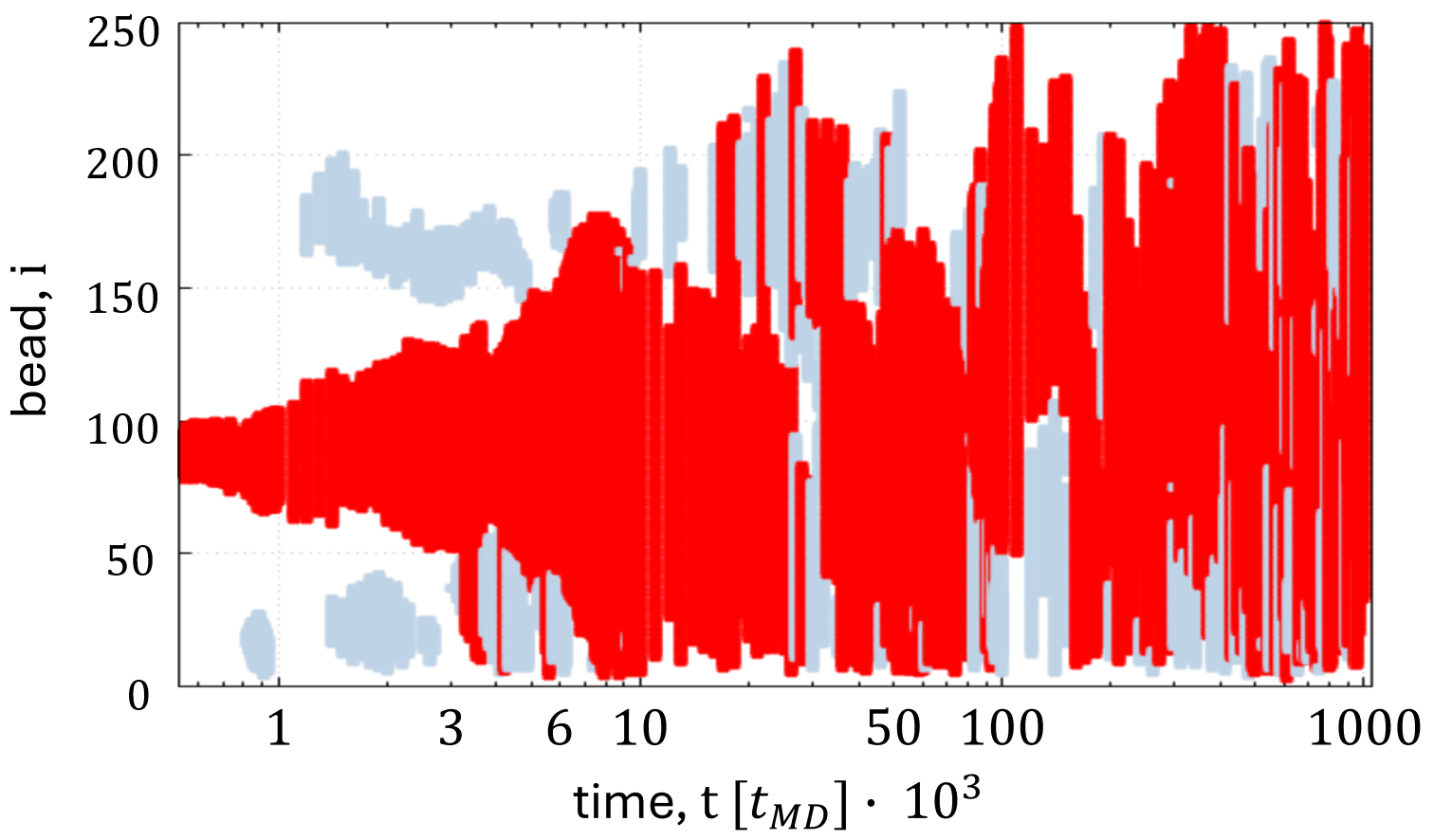}
    \caption{Kymograph of plectonemes positions, as a function of time, for the simulation of Fig.~\ref{fig:writhe_contact_map}, with $N=2 {\rm kbp}$ and $\tau_a=6$.
    The red regions correspond to the largest plectoneme, while the cyan ones indicate positions of additional plectonemes.}
    \label{fig:kymograph}
\end{figure}

Overall, Fig. \ref{fig:writhe_contact_map} highlights that a first plectoneme starts forming near the end where torque is applied, see peak in the local writhe around bead 80 and a larger off-diagonal band in the contact maps;  then the plectoneme grows over time until it eventually spans the entire chain. At intermediate times, we also observe the formation of additional smaller plectonemes, which merge with the larger one dynamically, similarly to what was observed in~\cite{Forte2019}.
This picture becomes clearer when examining the kymograph in Fig. \ref{fig:kymograph}, which plots regions of the chain with plectonemes, coloring the largest one in red and the smaller ones in cyan. The first plectoneme starts forming and growing at around $4l_b$ from the point of active torque application, while the others form immediately in the vicinity but do not grow and are eventually merged with the larger one.

\begin{figure}
    \centering
    \includegraphics[width=0.65\linewidth]{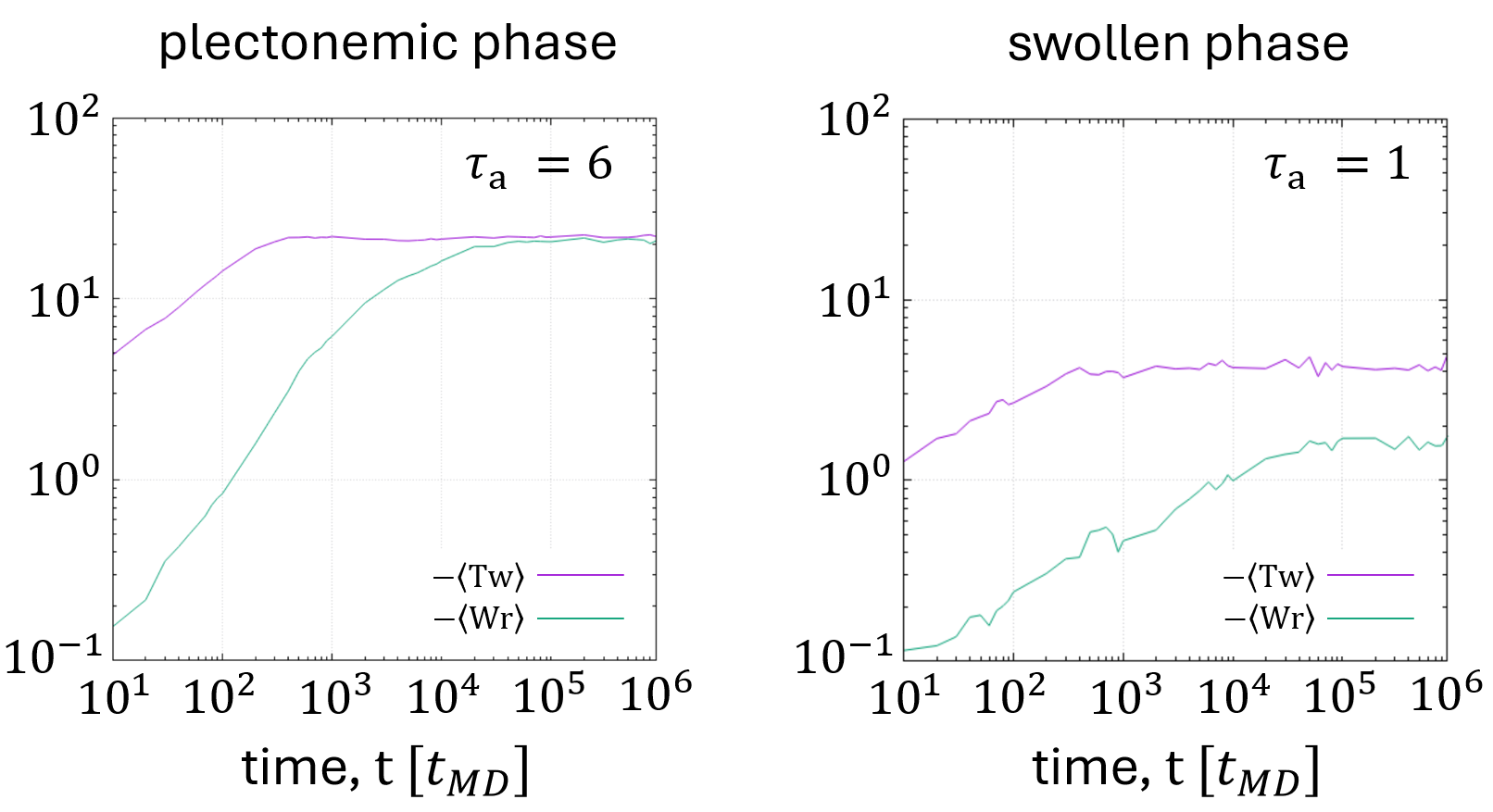}
    \caption{Total twist and writhe, {\rm Tw} and {\rm Wr} (Eqs.~\ref{totalTw} and \ref{eq:total_writhe}), as a function of time, for a $2$kbp long chain and $\tau_a=6$, averaged over several realizations. The values displayed in panels (a) and (b) are   
    for $\tau_a=6$ and $1$, corresponding to the plectonemic and swollen phases, respectively.}
    \label{fig:phases}
\end{figure}

Note that in our setup, at any given time, the sum of the total twist and writhe is not constrained to be constant. In fact, Fig.~\ref{fig:phases} shows {\rm Tw} and {\rm Wr}, averaged over different realizations, for $\tau_a=1, 6$ and a $2$kbp long chain. In both cases,  {\rm Tw} reaches a steady-state value faster than {\rm Wr}. Notably, at $\tau_a=1$, in the swollen phase, {\rm Tw} is much larger than  {\rm Wr}, while in the plectonemic phase at $\tau_a=6$ they reach a similar steady-state value.

\begin{figure}
    \centering
    \includegraphics[width=0.65\linewidth]{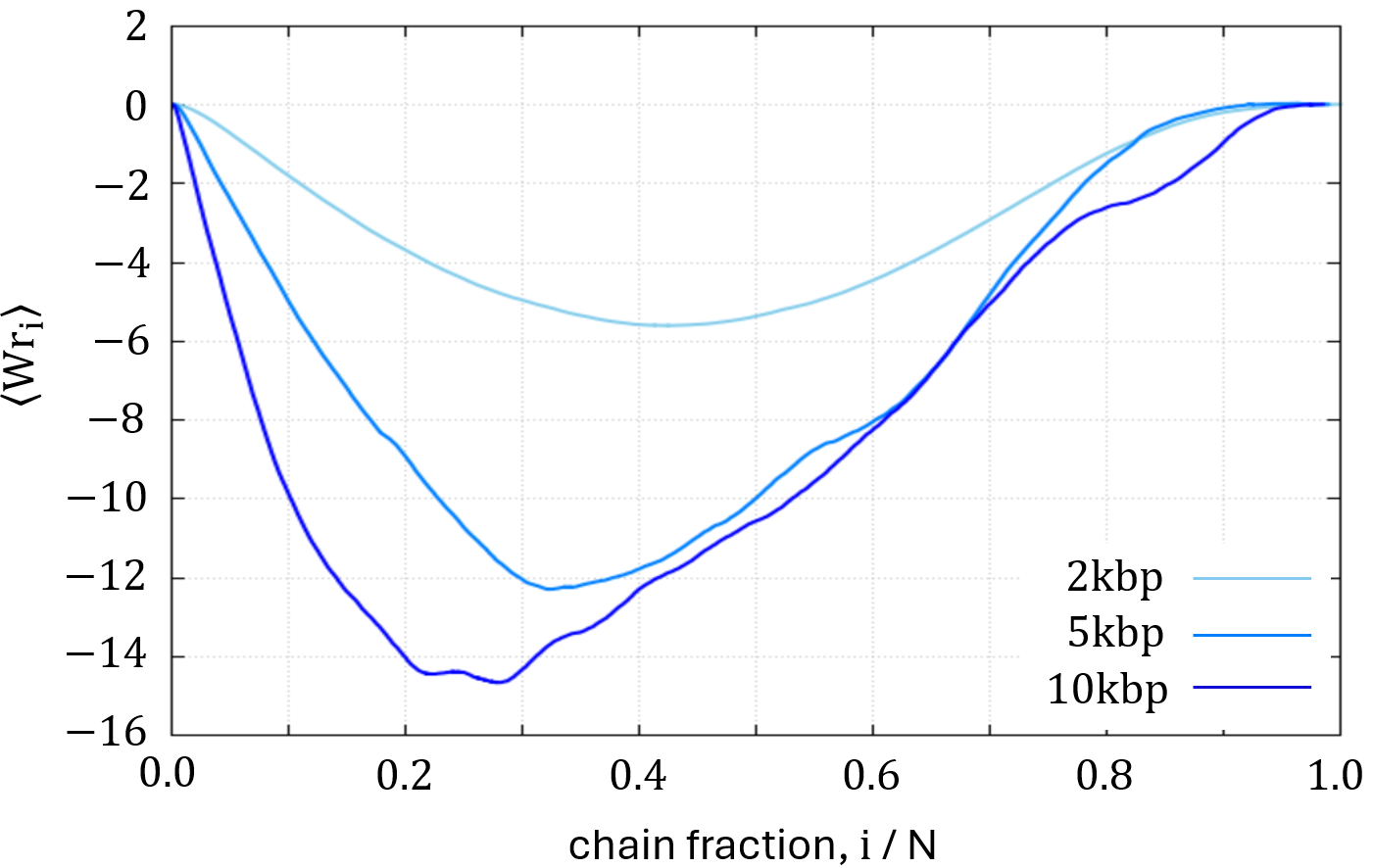}
    \caption{Local writhe ${\rm Wr_i}$ as a function of the chain fraction $i/N$, averaged over different conformations at steady state, shown for different chain lengths.}
    \label{fig:writhe_stat_chain}
\end{figure}

We can also characterize the local writhe profile, ${\rm Wr}_i$. Fig.~\ref{fig:writhe_stat_chain} shows these profiles, for $\tau_a=6$, averaged over different conformations at steady state.  We observe  that the curves are asymmetric. Increasing the chain length, we observe two phenomena: the peak position moves towards the chain end where torque is applied, while at the opposite end we have a small portion of the chain free from plectonemes. Thus, 
a striking feature is that, once formed, the torque effectively pins the plectoneme close to the point of active torque application.

Recent work by Skoruppa et al.~\cite{Skoruppa2022,Segers2025} has demonstrated the existence and statistical properties of multiplectoneme states in equilibrium, highlighting how chain length and torque control the number and distribution of these structures. 
In future work, it would be interesting to establish a more quantitative comparison between our driven scenario and the equilibrium predictions, particularly regarding the scaling of plectoneme length and the possibility of multiple domains emerging under different conditions.

\begin{figure}
    \centering
    \includegraphics[width=0.6\linewidth]{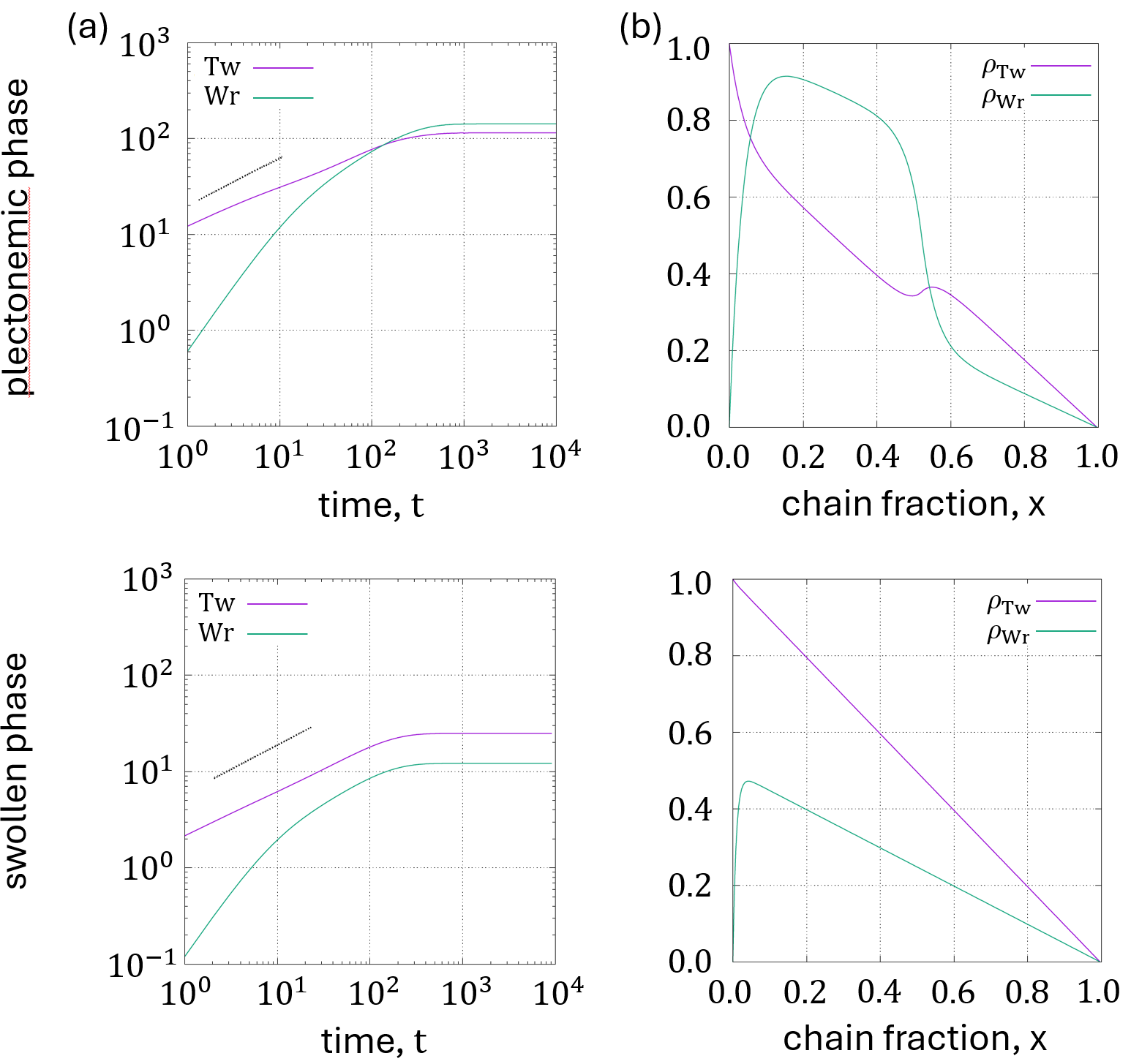} 
    \caption{The first and second rows show results from the statistical field theory, for the plectonemic and swollen phases, respectively. (a) Total twist and writhe, ${\rm Tw}$ and ${\rm Wr}$, as a function of time. The dashed line indicates $t^{0.5}$. 
    (b) Local twist density, $\rho_{\rm tw}$, and centerline writhe density,  $\rho_{\rm wr}$, as a function of the chain contour length $s$.  Parameters for the field theory model are: $\alpha=0.1$, $\beta=0.2$, $\gamma=0.15$, $\kappa=0.1$, and $\sigma_0=1$. For the swollen phase,  we set $D_{\rm tw}=100$, $D_{\rm wr}=1$, and $\rho_{{\rm tw}_0}=0.2$. For  the plectonemic phase we set $D_{\rm tw}=80$, $D_{\rm wr}=20$, and $\rho_{{\rm tw}_0}=1.2$.} 
    \label{fig:mean_field_theory}
\end{figure}

\textbf{Field theory for supercoiling transport.}
The qualitative trends observed up to now can be rationalized by using a field theory for supercoiling transport, which is based on the continuum equations of motion proposed in~\cite{Forte2019} for the coupled reaction-diffusion dynamics of local twist density,  $\rho_{\rm tw}(s,t)$, and centerline writhe density, $\rho_{\rm wr}(s,t)$, with $s$ the chain contour length, generalized to the boundary conditions which are appropriate for the setup we consider. These equations read as follows:
\begin{eqnarray}\label{meanfieldtheory}
    \frac{\partial\rho_{\rm tw}}{\partial t} & = & D_{\rm tw} \frac{\partial^2 \rho_{\rm tw}}{\partial s^2}-g(\rho_{\rm tw}, \rho_{\rm wr}) \\ \nonumber
    \frac{\partial\rho_{\rm wr}}{\partial t} & = & D_{\rm wr} \frac{\partial^2 \rho_{\rm wr}}{\partial s^2}-\kappa \frac{\partial^4 \rho_{\rm wr}}{\partial s^4} + g(\rho_{\rm tw}, \rho_{\rm wr}) \\ \nonumber
    g(\rho_{\rm tw}, \rho_{\rm wr}) & = & \alpha \rho_{\rm tw}-\beta\rho_{\rm wr}+\gamma \Theta(|\rho_{\rm tw}+\rho_{\rm wr}|-\sigma_0).
\end{eqnarray}
In Eqs.~(\ref{meanfieldtheory}), $D_{\rm tw}$ and $D_{\rm wr}$ denote respectively the twist and the writhe effective diffusion coefficients. At the same time, $\kappa$ is a constant which contributes to the surface tension of a plectoneme and the width of the interface between a plectoneme and a straight segment. 
Note that the equation of $\rho_{\rm tw}$ is of the same form as the one we found as Eq.\ref{eq:twist_evolution_continuous_cyl}, with the addition of the function $g(\rho_{\rm tw}, \rho_{\rm wr})$,  a reaction-like term which controls the mutual interconversion between local twist and local writhe. The active torque term in this case is imposed as a boundary condition. In the form we have chosen for $g$~\cite{Forte2019}, $\alpha$ and $\beta$ determine the rates of conversion from writhe to twist and vice versa, whereas $\gamma$ modulates this interconversion when the local supercoiling $\sigma_{sc}$ exceeds $\sigma_0$: if $\alpha\ge \beta \ge \gamma$, a chain with uniform twist is locally unstable to plectoneme formation as soon as $\sigma_{sc}>\sigma_0$. 
With respect to the field theory proposed in~\cite{Forte2019}, we disregard noise here -- this is reasonable, as we see little evidence of plectoneme diffusion along the chain in the simulations. Importantly, instead of periodic boundary conditions considered in~\cite{Forte2019}, here we consider Dirichlet-like boundary conditions for the twist and writhe fields, with $\rho_{\rm tw}(s=0)={\rho_{\rm tw}}_0$ at the end which is subject to the active torque, and $\rho_{\rm tw}(s=N\sigma)=0$ at the open end: these are equivalent to including a $\tau\delta'(x)$ forcing to the twist equation. For the writhe field, simulations suggest that $\rho_{\rm wr}$ should be $0$ at both ends. 
Note also that starting from Eq.~\ref{meanfieldtheory} one can obtain  an effective free energy  mathematically equivalent to the two-state free energy introduced by Marko~\cite{Marko2007}, see Ref.~\cite{Forte2019}.

We consider here two specific parameter sets. The first one, which we associate with the swollen phase, has $D_{\rm tw}/D_{\rm wr}=100$ and ${\rho_{\rm tw}}_0=0.2$ such that twist diffusion is much stronger than writhe diffusion, and activity is not strong enough to favor plectoneme formation. The second case, which we associate with the plectonemic phase, has $D_{\rm tw}/D_{\rm wr}=4$ and ${\rho_{\rm tw}}_0=1.2$  such that the diffusion of twist and writhe are compatible and plectoneme formation is favoured. This is reasonable, for instance, if plectonemes effectively slow down twist diffusion. Other simulation parameters are detailed in the caption of Fig.~\ref{fig:mean_field_theory}.

To compare with molecular dynamics simulations, we solve numerically Eqs.~(\ref{meanfieldtheory}) starting from a torsionally free initial condition with $\rho_{\rm tw}=\rho_{\rm wr}=0$. While the absence of noise and the phenomenological nature of our theory hinder a direct quantitative fit to the simulations, several trends are qualitatively consistent between the simulations and the mean field. 

Fig.~\ref{fig:mean_field_theory}(a) shows the total accumulated twist, ${\rm Tw}$ and writhe, ${\rm Wr}$ for the plectonemic and swollen cases, as a function of time, obtained integrating $\rho_{\rm tw}$ and $\rho_{\rm wr}$ over the entire chain length. First, in both cases, the twist accumulates over time approximately diffusively, as $t^{0.5}$, while the effective exponent of writhe accumulation varies over time. Twist and writhe reach a steady state eventually; the swollen and plectonemic phases are distinguishible in this model by the fact that in the first case the writhe is smaller than the twist, while in the second case they become of the same order. This results parallel the one of Fig.~\ref{fig:phases} for the polymer model.

Fig.~\ref{fig:mean_field_theory}(b) shows instead the steady-state profiles of $\rho_{\rm tw}$ and $\rho_{\rm wr}$ as a function of the contour length $s$. In the swollen phase, the local twist and writhe profiles decay linearly, with the former being more dominant and compatible with the expectation of Eq.\ref{eq:twist_distribution_cyl}. 
In the plectonemic phase, we observe instead a non-linear twist profile reminiscent of that found in molecular dynamic simulations, which corresponds to a slowdown in supercoiling transport due to plectoneme formation. 
The profile of $\rho_{\rm wr}$ shows again a linear decrease for $D_{\rm tw}\gg D_{\rm wr}$, while for $D_{\rm tw} \sim D_{\rm wr}$ a bump emerges, which should correspond to a region where the plectoneme establishes. Again, this region is adjacent to the point of active torque application, and parallels the one of Fig.~\ref{fig:writhe_stat_chain}, highlighting that the plectonemic phase is characterized by a writhe density contribution  dominant with respect to the twist density.
Note that in the plectonemic phase,  the mean field theory predicts also a small non-monotonic region in the twist profile close to the plectoneme boundary, which we do not found in simulations, possibly due to the neglect of noise in the field theory which disallows fluctuation in the position of such boundary, or due to the limited size of the chains considered in molecular dynamics simulations, as plectonemes span the entire chain.

\section*{Conclusions}
In summary, here we have studied a coarse-grained dynamical model for a twistable DNA molecule, in which each bead carries both positional and orientational degrees of freedom, allowing us to account for bending and twisting rigidities explicitly. By applying a constant torque at one of the two free chain ends, we probed the transport of torsional stress and the associated response across different torque magnitudes and chain lengths. This differs from typical setups, which consider a torsionally constrained molecule where, for instance, the ends are tethered. 

At sub-threshold torques ($\tau_a < \tau_{th}$, with $\tau_{th}\sim 2$), the chain is swollen and follows a behavior which is similar to that of a chain constrained to be straight: the intrinsic angular velocity is about constant along the chain and scale approximately as $\omega_L \sim \tau_a/N$, the local twist decays linearly along the chain, starting from the end where $\tau_a$ is applied, and the gyration radius does not change with respect to the value at $\tau_a=0$. 
Dynamically, the accumulation of total twist follows a diffusive law $\langle{\rm Tw} \rangle \sim t^{0.5}$ until saturating at the chain's torsional capacity, while the total bending remains mostly constant.

For $\tau_a \ge \tau_{th}$, there is a non-equilibrium elastic instability whereby local bending nucleates plectonemes, leading to a sharp reduction of the radius of gyration and a concomitant redistribution of torsional stress from twist into writhe. 
In this regime, the intrinsic angular velocity presents a slow decay, and concomitantly a global orbital rotation arises in the direction of $\tau_a$. The local twist becomes non-linear in the steady state, and a plectoneme appears near the driven boundary, paralleled by an increase in local bending. The fact that the local spin angular velocity decay in regions of high bend is concomitant with the emergence of a global rotation implies that these regions act as junctions where torque is distributed between the two rotational modes.
Dynamically, we still observe that twist diffuses initially, while multiple plectonemes form at the beginning and eventually fuse into a single one.  

The threshold value $\tau_{th}\sim 2$ (in units of $k_BT$) can be rationalised with a simple argument that balances the free energy cost of twist and bend. This thermodynamic argument assumes that energetic balances between twisting and bending still dominate the physics, even though the free boundary conditions we consider cause the chain to swivel in a steady state, placing the system out of thermodynamic equilibrium in this state. We highlight that the setup we consider differs from that analyzed in~\cite{Wada2009}, where a fixed rotational velocity ($\omega_0$, at the driven end) ensemble was considered instead of a fixed torque one, and where long-range hydrodynamic interactions were also considered. Additionally, we consider longer chains and reach a smaller angular velocity $\omega$ than in~\cite{Wada2009}. These differences are likely the reason why the scaling of $\omega$ with $\tau$ appears to be different from the observed $\omega\sim \omega_0^{1/3}$ scaling of~\cite{Wada2009}.

We have demonstrated that a reaction–diffusion mean-field theory for the coupled dynamics of local twist and writhe captures these trends qualitatively. 
In particular, the theory reproduces diffusive twist accumulation, linear twist profiles in the swollen regime, and a nonlinear steady-state twist profile in the plectonemic phase. The theory also shows that the non-linear twist profile arises due to interconversion into writhe, which is inhomogeneous along the chain and more pronounced near the driven boundary.

Taken together, these results demonstrate that there is a non-equilibrium torque-driven transition from a swollen phase with uniform torsional diffusion to localized supercoiling in the plectonemic phase, accompanied by an associated non-linear twist profile in the steady state. This transition is fundamentally distinct from the buckling of a tethered DNA under twist, which typically requires the presence of a stretching force~\cite{Neukirch2011,Marko1995}.  
The free-chain-end geometry that we have considered could be recreated in the lab through single-molecule experiments, for instance, by studying transcription on DNA templates in solution~\cite{Ma2014,Janissen2024}, via optical or magnetic tweezers~\cite{Strick1996,gao2021torsional}, or even via electrosmotic flow~\cite{zheng2025torsion,maffeo2023dna}.
Interestingly, the results in~\cite{Janissen2024} show that an RNA polymerase may be viewed as a torque-generated machine even if its rotational motion is not hindered. Because polymerases can exert a torque of a few $k_BT$~\cite{Ma2014,Jia2024}, it is possible that the torque-induced supercoiling with a free end we have studied is relevant in a biological context, where DNA or chromatin is transcribed~\cite{Liu1987}. Note that, of course, a transcribing polymerase would exert a torque {\it dipole} on the polymer~\cite{Liu1987}, resulting in the formation of twin domains. Therefore, our analysis in practice applies to the behaviour of one of the two supercoiling domains that would form. 

\section*{Conflicts of interest}
There are no conflicts to declare.

\section*{Acknowledgements}
D.Mo. is grateful to the Higgs Center for Theoretical Physics for financial support. G.F. acknowledges support from the Leverhulme Trust (Early Career Fellowship ECF-2024-221).

\end{document}